\documentclass[onecolumn,secnumarabic,amssymb, amsmath, nofootinbib,tightenlines,
nobibnotes, aps, prl,epsfig, showkeys]{revtex4}
\usepackage{graphicx}
\usepackage{dcolumn}
\usepackage{bm}

\usepackage{amssymb}
\usepackage{epsfig}
\usepackage{color}

\newcommand{\ba}{\begin{eqnarray}}
\newcommand{\ea}{\end{eqnarray}}
\newcommand{\be}{\begin{equation}}
\newcommand{\ee}{\end{equation}}
\newcommand{\bdisplay}{\begin{displaymath}}
\newcommand{\edisplay}{\end{displaymath}}

\begin{document}
\preprint{APS/123-QED}
\title{Transverse momentum dependent gluon density in a proton at low $x$ \\ in the Laplace transform method}

\author{G.R. Boroun}%
 \email{boroun@razi.ac.ir }
 \affiliation{Department of Physics, Razi University, Kermanshah
67149, Iran}%
\author{Phuoc Ha }
\email{pdha@towson.edu}
\affiliation{Department of Physics, Astronomy and Geosciences,
	Towson University, Towson, MD 21252}
\author{A.V. Kotikov }
\email{kotikov@theor.jinr.ru}
\affiliation{Bogolyubov Laboratory of Theoretical Physics, \\ Joint Institute for Nuclear Research,
Dubna 141980, Russia}%
\author{A.V. Lipatov }
\email{lipatov@theory.sinp.msu.ru}
\affiliation{Skobeltsyn Institute of Nuclear Physics, \\ Lomonosov Moscow State University, Moscow 119991, Russia}
\begin{abstract}
We investigate the gluon distribution in a proton at very low $x$, both integrated and transverse momentum dependent, using the Laplace transform technique. By accounting for leading and main next-to-leading contributions, we derive compact analytical expressions for the gluon densities valid in the asymptotic limit $x \to 0$. Our results closely match those from other analytical and numerical approaches, with the main advantage being the simplicity of the expressions, which capture the essential features of more complex calculations.


\end{abstract}
 \pacs{***}
\keywords{QCD evolution, small-$x$ physics, gluon densities in a proton} 

\maketitle


One of the key roles in the physical program
at modern and future colliders
is played by the Quantum Chromodynamics (QCD), 
since the main objects of research is the hard processes
occurring in $ep$ ($eA$) or $pp$ ($pA$) collisions. 
According to the QCD factorization hypothesis, the cross sections of such processes are calculated as a convolution of a hard scattering amplitude with the corresponding parton distribution functions (PDFs), which describe the momentum distributions of quarks and gluons inside hadrons.
For processes characterized by a single hard scale $\mu^2$, the collinear QCD factorization scheme is typically applied. In this approach, only the longitudinal momenta of the colliding partons are considered when calculating both the production amplitude and the PDFs, denoted as $a(x, \mu^2)$, where $a$ represents either a quark ($q$) or gluon ($g$).
The QCD evolution of the latter
is described by the Dokshitzer-Gribov-Lipatov-Altarelli-Parisi (DGLAP) equations \cite{DGLAP1, DGLAP2, DGLAP3, DGLAP4},
where large logarithmic terms proportional to $\alpha_s^n \ln^n \mu^2/\Lambda_{\rm QCD}^2$
are resummed.
In this framework, the transverse momenta of incoming partons ${\mathbf k}_T$ are neglected.
However, for a number of high energy processes involving several hard scales, the predictions of
collinear calculations become to be unreliable
since the interacting parton longitudinal momenta 
could be small and their transverse momenta cannot be neglected anymore.
This leads to the introduction of the
transverse momentum dependent parton distributions
(TMDs, or unintegrated parton densities, uPDFs) in the framework of high energy factorization \cite{HEF} (or $k_T$-factorization \cite{kT})
approach.
These quantities are more sensitive to parton dynamics than ordinary PDFs
and are one of the essential aspects of
modern high energy, or small $x$, physics (see reviews \cite{kT-review1, kT-review2} for more information.).

The TMD parton densities in a proton, $f_a(x, {\mathbf k}_T^2, \mu^2)$ are under active investigations at present.
Various models based on the different physical assumptions have been proposed recently (see, for example, \cite{TMDs-appr1, TMDs-appr2, TMDs-appr3, TMDs-appr4, TMDs-appr5} and references therein).
Usually, in analogy with ordinary PDFs, the initial TMD parton distributions should
be obtained first and then extended to a whole kinematic region using the non-collinear QCD evolution equations.
So, in the asymptotical $x \to 0$ limit, where
gluon contributions 
dominate over the quark component,
their density function obeys the Balitsky-Fadin-Kuraev-Lipatov (BFKL) evolution equation \cite{BFKL1, BFKL2, BFKL3}.
This equation resums large logarithmic terms proportional to $\alpha_s^n \ln^n s/\Lambda_{\rm QCD}^2 \sim \alpha_s^n \ln^n 1/x$.
An additional terms, proportional to $\alpha_s^n \ln^n 1/(1 - x)$ and essential at moderate and
large $x$, can be taken into account using the Catani-Ciafaloni-Fiorani-Marchesini (CCFM) equation \cite{CCFM}.
Moreover, non-linear effects, which could be important in the specific kinematical region,
are described by the Balitsky-Kovchegov (BK) \cite{BK} or JIMWLK \cite{JIMWLK1, JIMWLK2, JIMWLK3, JIMWLK4, JIMWLK5, JIMWLK6} equations.
There are also several approaches 
based on conventional
DGLAP scenarios, namely, recent Parton Branching (PB) method \cite{PB1, PB2} and
Kimber-Martin-Ryskin (KMR) \cite{KMR}, or Watt-Martin-Ryskin (WMR) \cite{WMR}
prescriptions, where
ordinary PDFs derived from the numerical or analytical solutions
of the DGLAP evolution equations are used as an input (see also \cite{KWMR-discussion1, KWMR-discussion2, KWMR-discussion3, KWMR-discussion4}).
Nevertheless, the proton TMDs 
still remain to be
rather poorly known quantities in comparison with their conventional (collinear) counterpart, PDFs.

This work is dedicated to further investigation of the gluon dynamics in a proton.
We perform the analytical solution of the DGLAP equations for ordinary PDFs
with taking into account leading order (LO) and main next-to-leading order (NLO) contributions at low $x$ and then
derive corresponding expressions for TMDs.
Our method is based on the Laplace-transform 
technique\footnote{The Mellin transform, which is 
more popular in the literature on high-energy physics, 
in the variable $x$ after the restriction $x\leq 1$ is exactly equal to the Laplace transform in the variable $\upsilon$.
Without such restriction, the Mellin transform corresponds to the so-called two-sided Laplace transform (see \cite{Mellin}).} \cite{Block1, Block2, Block3, Block4, BH1, BH2},
which has been succesfully applied earlier, in particular, to study
proton longitudinal structure function $F_L(x, Q^2)$ \cite{BH1}.
The importance of such investigations lies in the fact that
the detailed knowledge of TMDs is essential for
theoretical predictions for number of multiscale QCD processes studied at modern and future colliders. 
Of course, it has impact on present
measurements at the LHC and influences the preparation for measurements at LHeC, EIC, EicC, NICA and CEPC.
Main advantage of our approach is related with
simple analytical expressions derived for both ordinary and
TMD gluon densities in a proton which could reproduce the
main features of rather complicated numerical calculations.





We start from well-known fact that,
at very low $x$, the TMD gluon distribution
becomes independent on the hard scale $\mu^2$.
In this kinematical region, the approximated (and often used) relation with ordinary gluon density could be valid:
\ba \label{UGD_eq}
  f_g(x,{\mathbf k}_{T}^{2})\simeq \left. \frac{\partial}{{\partial} \mu^{2}} \, G(x,\mu^{2})\right\vert_{\mu^2={\mathbf k}_{T}^{2}},
\ea
\noindent
where $G(x,\mu^2) \equiv xg(x,\mu^2)$ and transverse momentum components along the evolution ladder are the same order in the BFKL kinematics \cite{BFKL1, BFKL2, BFKL3}.
Ignoring the quark contributions in the asymptotical limit $x \to 0$,
the DGLAP evolution equation for the gluon distribution function
reads
%
%
%
%
\ba \label{DGLAP1_eq}
\frac{\partial}{\partial{\ln{\mu^2}}}\,{G}(x,\mu^2) {\simeq} \int\limits_{x}^{1}P_{gg}\left(z,\alpha_{s}\left(\mu^2\right)\right)G\left(\frac{x}{z},\mu^2\right)dz,
\ea
\noindent
where
the splitting function $P_{gg}\left(z,\alpha_{s}\left(\mu^2\right)\right)$ 
is given by the following form
\ba
\label{Coupling_eq}
P_{gg}\left(z,\alpha_s\left(\mu^2\right)\right) = \sum_{n=1} \left(\frac{\alpha_{s}\left(\mu^2\right)}{4\pi}\right)^{n} P^{(n)}_{gg}(z) = \sum_{n=1} \left(a_{s}\left(\mu^2\right)\right)^{n} P^{(n)}_{gg}(z).
\ea
\noindent
Here we denote $a_s(\mu^2) = \alpha_{s}\left(\mu^2\right)/4\pi$.
Using the notation
$\widehat{G}(\upsilon, \mu^2) \equiv G(e^{-\upsilon}, \mu^2)$, we
can rewrite the DGLAP equation for the gluon distribution in
terms of the variables $\upsilon={\ln}(1/x)$ and $t = \ln \mu^2$:
\ba \label{DGLAP2_eq}
\frac{\partial}{\partial t} \, {\widehat{G}}(\upsilon, t) = \int\limits_{0}^{\upsilon}\widehat{H}^{(n)}_{gg}\left(\upsilon-w,a_{s}\left(t\right)\right)\widehat{G} \left(w,t\right)dw,
\ea
\noindent
where $\widehat{H}^{(n)}_{gg}\left(\upsilon,a_{s}\left( t\right)\right) = e^{-\upsilon}\widehat{P}_{gg}\left(\upsilon,a_{s}\left( t \right)\right)$.
By introducing appropriate notations and applying Laplace transform
techniques \cite{Block1, Block2, Block3, Block4, BH1, BH2}, we can
express the solution of the evolution equation~(\ref{DGLAP2_eq}) in terms of the Laplace
transforms of the relevant factors:
\ba \label{Laplace1_eq}
\frac{\partial}{\partial t} \, {g}(s, t) = \sum_{n=1}h^{(n)}_{gg}\left(s,a_{s}\left( t \right)\right) g\left(s, t\right),
\ea
\noindent
where the Laplace transform of
$\widehat{H}^{(n)}_{gg}\left(\upsilon,a_{s}\left( t \right)\right)$ is given by
\ba \label{Laplace1a_eq}
h^{(n)}_{gg}\left(s,a_{s}\left( t \right)\right) \equiv \mathcal{L}\left[\widehat{H}^{(n)}_{gg}\left(\upsilon,a_{s}\left( t \right)\right);s\right] = \int\limits_{0}^{\infty} \widehat{H}^{(n)}_{gg}\left(\upsilon,a_{s}\left( t \right)\right)e^{-s\upsilon}d\upsilon.
\ea
\noindent
Using~(\ref{Laplace1_eq}) and (\ref{Laplace1a_eq}), we can easily obtain
\begin{gather}
{g}(s, t) = {g}_{0}(s, t_0) \, \exp \left[{ \, \, \, \int\limits_{t_{0}}^t} \sum_{n=1}h^{(n)}_{gg}\left(s,a_{s}\left(t^\prime\right)\right)d{t^\prime}\right] =
{g}_{0}(s, t_0)\, \exp \left[ \, \sum_{n=1}h^{(n)}_{gg}(s){\int\limits_{t_{0}}^{t}}a^{n}_{s}(t^\prime)d {t^\prime}\right] \simeq \nonumber \\
 \simeq g_0(s,t_0) + {g}_{0}(s,t_0) \left[h^{(1)}_{gg}(s){\int\limits_{t_0}^t}a_{s}(t^\prime)dt^\prime + h^{(2)}_{gg}(s)\int\limits_{t_0}^{t} a^{2}_{s}(t^\prime)d{t^\prime}+\frac{1}{2}\left(h^{(1)}_{gg}(s){\int\limits_{t_{0}}^{t}}a_{s}(t^\prime)d{t^\prime}\right)^2 \, \, \, \right] + \mathcal{O}(a_{s}^{3}),
     \label{Laplace2_eq}
\end{gather}
\noindent
where ${g}_{0}(s,t_0)$ is the non-perturbative initial gluon density determined at some starting scale $t_0 \equiv \ln \mu_0^2$ and we ignored $\mathcal{O}(a_{s}^{3})$ terms and higher.
The leading order (LO) coefficient $h^{(1)}_{gg}(s)$ is given by
\ba \label{Coef1_eq}
h^{(1)}_{gg}(s) = \frac{33-2n_{f}}{3}+12\left[\frac{1}{s}-\frac{2}{s+1} + \frac{1}{s+2}-\frac{1}{s+3}-\psi(s+1)-\gamma_{E} \right],
\ea
\noindent
where $\psi(s)$ is the digamma function, $\gamma_{E} \simeq 0.5772156$ is Euler's constant
and $n_f$ is the number of active quark flavors.
The evaluation of the next-to-leading (NLO) coefficients is straightforward, but too lengthy to be included in this note.
Keeping only the largest terms in the limit $s \to 0$, we have
\ba \label{Coef2_eq}
\left. h^{(2)}(s)\right\vert_{s \to 0} \simeq {1\over s} \left(\frac{4}{3}C_{F}T_{f} - \frac{46}{9}C_{A}T_{f}\right) - {1\over s^2} \, 4 C^{2}_{A} \ln 2
\ea
\noindent
with $C_{F}= (N_{c}^2-1)/2N_{c}$, $C_{A}=N_{c}$, $T_{R}=1/2$, $T_{f}=T_{R}n_{f}$.

The perturbative expansion of the exponent of the renormalization group according to~(\ref{Laplace2_eq})
leads to a possibility to obtain accurate analytical results 
using the Laplace transform technique \cite{Block1, Block2, Block3, Block4, BH1, BH2}. Of course, 
this expansion restricts the use of $t$ values to be close to $t_0$. 
However, the restriction is not very strong\footnote{There is also another approach \cite{Mellin_ap} based on the Mellin transform,
which exactly preserves the exponent of the renormalization group. However, its application is limited
only by the low $x$ range and flat ($a = 0$) initial conditions, see below.} 
for hard scale $\mu^2$.
Then, in accordance with~(\ref{UGD_eq}) and~(\ref{Laplace2_eq}), the TMD gluon density in $s$-space can be easily
rewritten as:
\begin{gather}
  \left. f_g(s,t) \right\vert_{t \geq {t_0}} = {g}_{0}(s,t_0)\frac{\partial}{{\partial}t}\left[h^{(1)}_{gg}(s){\int\limits_{t_{0}}^{t}}a_{s}(t^\prime)d{t} + h^{(2)}_{gg}(s){\int\limits_{t_{0}}^{t}}a^{2}_{s}(t^\prime)d{t^\prime}
+\frac{1}{2}\left(h^{(1)}_{gg}(s){\int\limits_{t_{0}}^{t}}a_{s}(t^\prime)d{t^\prime}\right)^2 \, \, \, \right],
 \label{UGDs_eq}
\end{gather}
\noindent
where variable $t = \ln {\mathbf k}_T^2 \geq t_0 \equiv \ln k_0^2$ is now related with the gluon transverse momentum ${\mathbf k}_T^2$.
Of course, in the infrared region, $t < t_0$, the TMD gluon density have to be modelled.
Following \cite{Jung}, we assume that, in $x$-space, it could be parametrized in a rather general form, namely,
\ba
  \left. f_0(x, {\mathbf k}_T^2)\right\vert_{{\mathbf k}_T^2 < k_0^2} = N x^{a}(1-x)^{b} \exp \left( - {\mathbf k}_T^2/k_0^2 \right),
   \label{UGDini_eq}
\ea
\noindent
where $N$, $a$, $b$ and $k_0$ are the free parameters.
Note that ordinary gluon density $g_0(x, \mu_0^2)$ can be obtained
by the integration of~(\ref{UGDini_eq}) over ${\mathbf k}_T^2$ up to $\mu_0^2$.

Using the general expression~(\ref{UGDini_eq}) for initial gluon distribution,
%
%
%
%
the inverse Laplace transform of~(\ref{Laplace2_eq}) and/or (\ref{UGDs_eq}) 
is straightforward and could be performed analytically 
in the high-energy region of the coefficients $h^{(n)}_{gg}(s)$.
Here we employ the method developed \cite{Block1, Block2, Block3, Block4, BH1, BH2}.
In fact,
let $F(s)$ be the Laplace transform of a function
$f(\upsilon)$. It is straightforward to show that the inverse
Laplace transform of $F(s)/s$ is $\int\limits_0^{\upsilon} du f(u)$ and
inverse Laplace transform of $F(s)/s^2$ is $\int\limits_0^{\upsilon}
du \int\limits_0^u dwf(w)$. For $f(\upsilon) \sim
(e^{-\upsilon})^a(1-e^{-\upsilon})^b$, that corresponds to~(\ref{UGDini_eq}), we find that the Laplace
transform of $f(\upsilon)$ is
\ba \label{F1_eq}
 F(s) = \mathcal{L} \left[f(\upsilon); s \right] \sim \mathcal{L} \left[e^{-a\upsilon}(1-e^{-\upsilon})^b; s \right] = \frac{\Gamma(a+s) \Gamma(1+b)}{\Gamma(1+ a+b+s)}.
\ea
\noindent
It follows that  the inverse Laplace transforms of $F(s)/s$ and
$F(s)/s^2$ are given by
\begin{gather} \label{F2_eq}
 \mathcal{L}^{-1} \left[\frac{F(s)}{s}; \upsilon \right] = \int\limits_0^{\upsilon} f(u) du \sim  B(1,a, 1+b) - B(e^{-\upsilon}, a, 1+b), \\
 \label{invFs2 eq}
\mathcal{L}^{-1} \left[\frac{F(s)}{s^2}; \upsilon \right] = \int\limits_0^{\upsilon} du \int\limits_0^{u} f(w) dw \sim  B(1,a, 1+b) \, \upsilon -  \int\limits_0^{\upsilon} dw \, B(e^{-w}, a, 1+b) ,
\end{gather}
\noindent
where $B(z,a,b)$ is the incomplete Beta function. The last integral in~(\ref{invFs2 eq}) is found to be
\ba \label{F3_eq}
 \int\limits_0^{\upsilon} dw \, B(e^{-w}, a, 1+b) = - \frac{e^{-a \upsilon}}{a^2} \, _{3}F_{2}(a, a ,-b; 1+a, 1+a; e^{- \upsilon}) + \frac{\Gamma(a) \Gamma(1+b)}{\Gamma(1+ a+b)} \left[\psi (1+a+b) - \psi (a) \right],
\ea
\noindent
where $_{3}F_{2}(a_1,a_2,a_3; b_1, b_2; z)$ is a generalized hypergeometric function\footnote{For $xg_0(x,\mu_0^2) \sim x^a (1-x)^b$, the 
expressions~(\ref{invFs2 eq}) and (\ref{F3_eq}) can be obtained also as
$\mathcal{L}^{-1} [F(s)/s^2];\,\upsilon]=(d/da + \upsilon) \mathcal{L}^{-1} [F(s)/s;\,\upsilon ]$.}.
%
%
In the asymptotical limit $s \to 0$, we retain only the $1/s$ and
$1/s^2$ terms in the expressions of $h^{(1)}_{gg}(s)$ and
$h^{(2)}_{gg}(s)$. Using the results~(\ref{F1_eq}) --- (\ref{F3_eq}) to
calculate the inverse Laplace transform of~(\ref{UGDs_eq}) and then
returning to $x$-space, we get
%
%
%
%
%
%
\begin{gather}
 xg(x,\mu^2) = xg_{0}(x,\mu_0^2) + g_1(\mu_0^2) \left[f_{1}(x){\int\limits_{t_{0}}^{t}} a_{s}(t^\prime)d{t^\prime} + f_{2}(x){\int\limits_{t_{0}}^{t}} a^{2}_{s}(t^\prime)d{t^\prime} + \frac{1}{2}f_{3}(x) \left(\, \, \, \int\limits_{t_{0}}^{t} a_{s}(t^\prime)d{t^\prime}\right)^2 \, \, \, \right],
  \label{GDx_eq}
\end{gather}
\begin{gather}
  \left. f_g(x, {\mathbf k}_T^2)\right\vert_{{\mathbf k}_T^2 \geq k_0^2} = {g_1(k_0^2) \over {\mathbf k}_T^2} \left[ a_{s}\left({\mathbf k}_T^2\right) f_{1}(x) + a_{s}^2\left({\mathbf k}_T^2\right) f_{2}(x) + a_{s}\left({\mathbf k}_T^2\right) f_{3}(x) \int\limits_{t_{0}}^{t} a_{s}(t^\prime)d{t^\prime} \right],
  \label{UGDx_eq}
\end{gather}
\noindent
where the gluon density $f_g(x, {\mathbf k}_T^2)$ at low ${\mathbf k}_T^2 < k_0^2$ is given by~(\ref{UGDini_eq}) and
\begin{gather}
  f_{1}(x) = 12 \left[  B(1,a, 1+b) - B(x, a, 1+b) \right], \quad f_{2}(x) = -\frac{61}{27}f_{1}(x)  - \frac{ \ln 2}{4} f_{3}(x), \nonumber \\
  f_{3}(x) = 144 \Bigg[ B(1,a, 1+b) \, \ln 1/x  +  \frac{x^{a}}{a^2} \, _{3}F_{2}(a, a ,-b; 1+a, 1+a; x) - \frac{\Gamma(a) \Gamma(1+b)}{\Gamma(1+ a+b)} \left[ \psi(1+a+b) - \psi (a) \right] \Bigg], \nonumber \\
  xg_0(x,\mu_0^2) = x^a (1 - x)^b g_1(\mu_0^2), \quad g_1(\mu^2) = N k_0^2 \left[ 1 -  \exp \left( - \mu^2/k_0^2 \right) \right].
\label{f0x_eq}
\end{gather}
\noindent
Here we use the approximation $h_{gg}^{(1)}=12/s$, which gives the leading contribution at $x\to 0$.
Note that variable $t$ involved to~(\ref{UGDx_eq}) is related with the gluon transverse momentum ${\mathbf k}_T^2$.
This is in contrast with (\ref{GDx_eq}), where it is related with the hard scale $\mu^2$.
Moreover, when we limit ourselves to a low $x$ range, we can expand the the incomplete
Beta function and hypergeometric function with respect of $x$, preserving the asymptotic term and the
first subasymptotic term: $B(x, a, 1+b)\approx x^a (1/a-bx/(1+a))$ and $_{3}F_{2}(a, a ,-b; 1+a, 1+a; x)
\approx 1-a^2bx/(1+a)^2$. The subasymptotic term $\sim x$ shows us the accuracy of our analysis. So,
the results for $f_1(x)$ and $f_3(x)$ are simplified.

\begin{figure}
    \begin{center}
        \includegraphics[width=5.9cm]{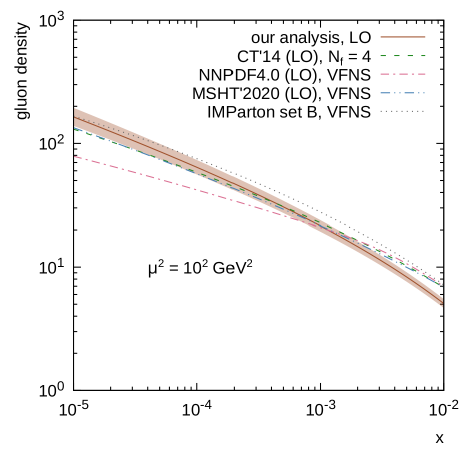}
        \includegraphics[width=5.9cm]{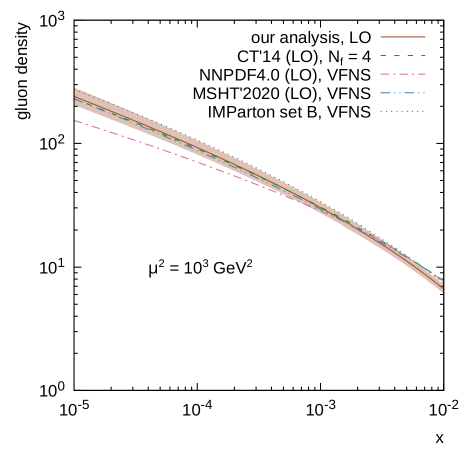}
        \includegraphics[width=5.9cm]{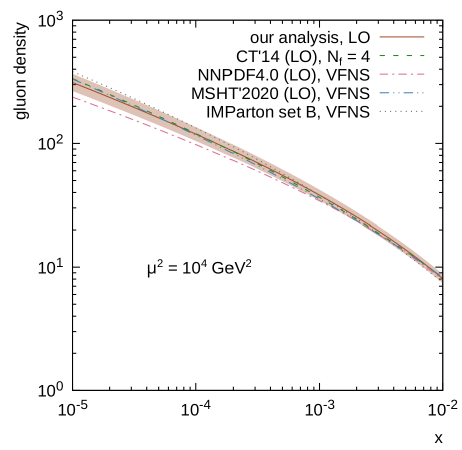}
        \caption{Conventional gluon density in a proton $xg(x,\mu^2)$ calculated at the LO as function of $x$ for different
            values of $\mu^2$. For comparison we show here the results of numerical solutions of the LO DGLAP
            equations performed by the CTEQ-TEA \cite{CT14}, NNPDF4.0 \cite{NNPDF4}, MSHT'2020 \cite{MSHT20} and IMP \cite{IMP} groups.}
        \label{fig:1}
    \end{center}
\end{figure}

\begin{figure}
    \begin{center}
        \includegraphics[width=5.9cm]{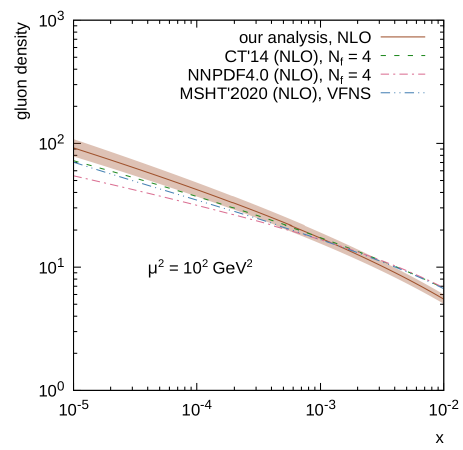}
        \includegraphics[width=5.9cm]{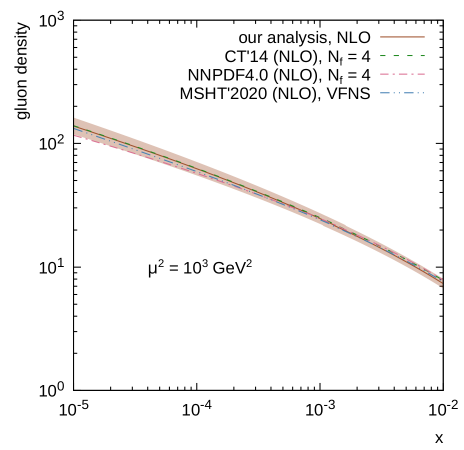}
        \includegraphics[width=5.9cm]{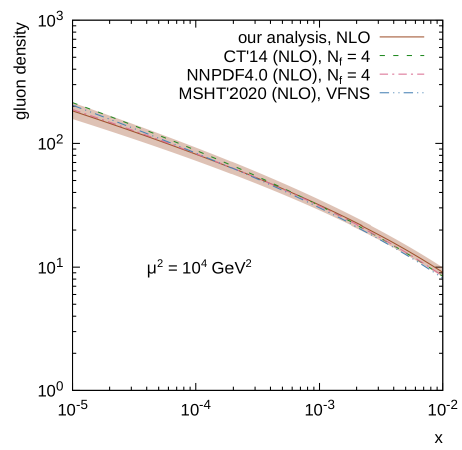}
        \caption{Conventional gluon density in a proton $xg(x,\mu^2)$ calculated at the NLO as function of $x$ for different
            values of $\mu^2$. For comparison we show here the results of numerical solutions of the NLO DGLAP
            equations performed by the CTEQ-TEA \cite{CT18}, NNPDF4.0 \cite{NNPDF4} and MSHT'2020 \cite{MSHT20} groups.}
        \label{fig:2}
    \end{center}
\end{figure}

\begin{figure}
    \begin{center}
        \includegraphics[width=5.9cm]{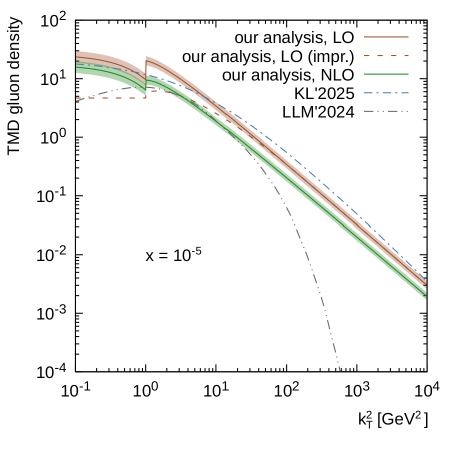}
        \includegraphics[width=5.9cm]{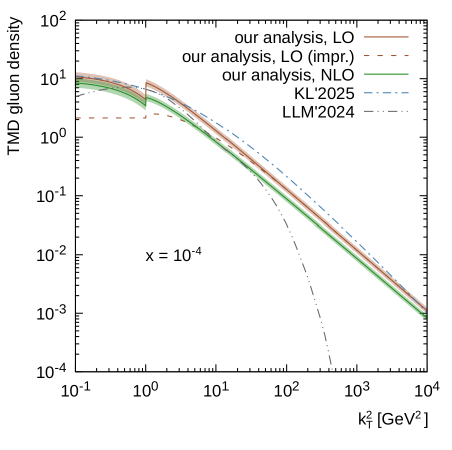}
        \includegraphics[width=5.9cm]{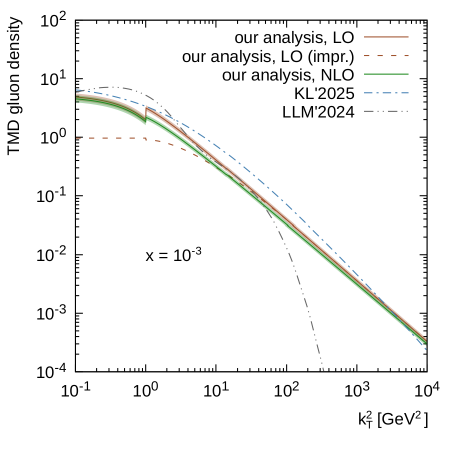}
        \caption{TMD gluon densities in a proton $f_g(x, {\mathbf k}_T^2)$ calculated as function of ${\mathbf k}_T^2$
            for different values of $x$. For comparison we show corresponding results obtained within the KMR/WMR and CCFM approaches, labeled as KL'2025 \cite{TMDs-appr5} and LLM'2024 \cite{TMDs-appr2}, respectively. Note that we set $\mu^2 = 100$~GeV$^2$ for two-scale involved gluon distributions KL'2025, LLM'2024 and the one given by~(\ref{eq-improved-gluon}).}
        \label{fig:3}
    \end{center}
\end{figure}

The simple analytical expressions above could be easily used to investigate the small-$x$ asymptotics
of the gluon distribution functions in a proton.
As an illustration, in Figs.~\ref{fig:1} and \ref{fig:2} we show
the collinear gluon densities $xg(x,\mu^2)$ derived at the LO and NLO
according to~(\ref{GDx_eq}) as a functions of $x$ for several $\mu^2$ values at $x \leq 10^{-2}$.
Here we set $a = - 0.34 \pm 0.02$, $b = 12.0 \pm 1.0$, $N = 0.52$~GeV$^{-2}$
in the LO calculations and $a = - 0.27 \pm 0.02$, $b = 6.0 \pm 1.0$, $N = 0.78$~GeV$^{-2}$
in the NLO ones. Our choice for $a$ and $b$ values is similar to that in \cite{Jung}.
The shaded bands represent the estimation of the uncertainties
connected with some reasonable variations in the $a$ and $b$ parameters
added in quadratures.
For comparison, we also plot corresponding
results obtained
in the fixed-flavor-number-sheme (FFNS) 
and/or variable-flavor-number-sheme (VFNS)
by the CTEQ-TEA \cite{CT14, CT18}, NNPDF4.0 \cite{NNPDF4}, MSHT'2020 \cite{MSHT20} and IMP \cite{IMP} groups.
As one can see, the reasonable well agreement between our analytical LO and NLO results and relevant numerical analyses
could be achieved for moderate and large $\mu^2$
with appropriate choice for phenomenological parameters of input gluon density (\ref{UGDini_eq}).
In particular, our fit and FFNS predictions from CTEQ-TEA Collaboration are very close to each other.
At lower $\mu^2$ values, some discrepancy between the different PDFs could be observed,
that is mainly due to essentially different initial conditions used in the considered analyses.
Note that we have applied a one (two) loop formula for QCD strong coupling with $N_f = 4$ and
$\Lambda_{\rm QCD} = 200$~MeV in the LO (NLO) computations.
Now we turn to the TMD gluon density in a proton. In
Fig.~\ref{fig:3} we show $f_g(x,{\mathbf k}_T^2)$ calculated at
the LO and NLO as function of transverse momentum ${\mathbf
k}_T^2$ for several values of $x$. Shaded bands represent the
uncertainties connected with the variation in the $a$ and $b$
parameters as it was indicated above. Some discontinuity observed
at ${\mathbf k}_T^2 = k_0^2 = 1$~GeV$^2$ is related with the
general normalization condition (\ref{UGD_eq}) which holds after
splitting of the $f_g(x,{\mathbf k}_T^2)$ into the modelled soft
and perturbative parts given by (\ref{UGDini_eq}) and
(\ref{UGDx_eq}), respectively. Our results are compared with
predictions \cite{TMDs-appr5} obtained in the KMR/WMR approach
\cite{KMR,WMR} and scenario \cite{TMDs-appr2} based on the CCFM
evolution \cite{CCFM}. First of them (KL'2025) was derived at
leading order of the QCD running coupling, where analytical
solution of DGLAP equations for conventional PDFs, valid in a wide
kinematical region, has been used as an input. The second
(LLM'2024) gluon density is based on the CCFM-evolved\footnote{In
the leading logarithmic approximation, see \cite{CCFM, Jung} for
more details.} expression for initial gluon distribution which
provides a self-consistent simultaneous description of number of
the collider data, in particular, HERA data on the electron-proton
deep inelastic scattering and soft hadron production in $pp$
collisions at the LHC conditions. Of course, all these approaches
represent the essentially different ways to evaluate the TMD gluon
density in a proton. However, we find a notable agreement of our
LO results and LLM'2024 predictions at moderate gluon transverse
momenta, ${\mathbf k}_T^2 \leq 100$~GeV$^2$ in a wide $x$ range,
as it is clearly demonstrated in Fig.~\ref{fig:3}. It indicates
again the applicability of developed Laplace transform technique
\cite{Block1, Block2, Block3, Block4, BH1, BH2} to the small-$x$
QCD processes. The observed agreement between our simple formulas
and more complicated considerations \cite{TMDs-appr5, TMDs-appr2} 
is very promising for forthcoming and more detailed investigations.
At the same time, different behavior of gluon densities at low ${\mathbf k}_T^2$
is due to different assumptions applied in the soft region, ${\mathbf k}_T^2 < k_0^2$,
while discrepancies at very large transverse momenta
are related with the distinctions in the QCD evolution.

The consideration could be improved by taking into account 
Sudakov form factor (see \cite{KMR, WMR}). Then, the main formula~(\ref{UGD_eq}) is rewritten in the form:
\begin{gather} 
f_g(x,{\mathbf k}_{T}^{2}, \mu^2)\simeq \frac{\partial}{{\partial} {\mathbf k}_T^{2}} \left[ T_g({\mathbf k}_T^{2}, \mu^2) \, G(x,{\mathbf k}_T^{2}) \right],
\label{eq-improved-gluon}
\end{gather}
\noindent
where TMD gluon density becomes two-scale dependent and Sudakov form factor $T_g({\mathbf k}_T^{2}, \mu^2)$ resums the virtual 
contributions up to all orders\footnote{In the previos studies \cite{TMDs-appr5} we used the 
numerical results for $T_g({\mathbf k}_T^2, \mu^2)$. Here we use both the numerical and analytic results.}:
\begin{gather}
	\ln T_g({\mathbf k}_T^2, \mu^2) = - 2\int\limits_{{\mathbf k}_T^2}^{\mu^2} {d {\mathbf p}_T^2 \over {\mathbf p}_T^2} a_s({\mathbf p}_T^2)
	\sum_q \int\limits_0^{1 - \Delta} dz z P_{qg}^{(1)}(z),
	\label{eq-sudakov}
\end{gather}
\noindent
with $P_{ab}^{(1)}(z)$ being the LO splitting functions.
From~(\ref{eq-improved-gluon}) we can easily obtain
\begin{gather}
\left. f_g(x,{\mathbf k}_{T}^{2}, \mu^2)\right\vert_{{\mathbf k}_T^2 \geq k_0^2} = 4C_A \frac{a_s({\mathbf k}_T^2)}{{\mathbf k}_T^2}\,R_g(\Delta)
T_g({\mathbf k}_T^{2}, \mu^2) \, xg(x,{\mathbf k}_T^{2}) + T_g({\mathbf k}_T^{2}, \mu^2) \, f_g(x,{\mathbf k}_T^{2}),
\label{eq-improved-gluon1}
\end{gather}
where $xg(x,{\mathbf k}_T^{2})$ and $f_g(x,{\mathbf k}_T^{2})$ are given by (\ref{GDx_eq}) and (\ref{UGDx_eq}), respectively, and
\begin{gather}
R_g(\Delta)=\frac{1}{2C_A} \,\sum_q \int\limits_0^{1 - \Delta} dz z P_{qg}^{(1)}(z).
\label{Rg}
\end{gather}
The analytic results for $R_g(\Delta)$ and 
$T_g({\mathbf k}_T^{2}, \mu^2)$ are given in Appendix.
The Sudakov form factor describes the probability that a quark
propagating through a vacuum will not radiate any soft gluons and
vanishes exponentially at large energy.
Usually, the cut-off parameter $\Delta$ 
is choosen to be $\Delta = |{\mathbf k}_T|/\mu$ or $\Delta = |{\mathbf k}_T|/(|{\mathbf k}_T| + \mu)$,
that reflects the strong ordering or angular ordering conditions for parton
emissions at the last evolution step to regulate the soft gluon singularities (see \cite{KMR,WMR} and 
discussions \cite{KWMR-discussion1, KWMR-discussion2, KWMR-discussion3, KWMR-discussion4}).
The effect of the
Sudakov form factors in the TMD gluon distribution functions are shown in Fig.~\ref{fig:3}
for the LO approximation.
Differences between the results (with and without Sudakov form factor) 
are small and visible at small transverse momenta.
The Sudakov factor mostly disappears in the large-${\mathbf k}_T^2$
region. The TMD gluon density with Sudakov form factor seem to
show slightly more change in a wide range of ${\mathbf k}_T^2$.
Note that the plateu at ${\mathbf k}_T^2 < k_0^2$ is due to 
modelled gluon density given by~(\ref{eq-improved-gluon1}) in the soft region, $f_g(x, {\mathbf k}_T^2, \mu^2) = xg(x,k_0^2) T_g(k_0^2, \mu^2)/k_0^2$ \cite{KMR, WMR}.
Of course, the inspection of corresponding phenomenological consequences
needs the dedicated study and out of our present short note.

To conclude,
we have considered the behavior of ordinary and transverse momentum dependent
gluon distribution functions in a proton at low $x$.
We have developed a method for the analytical derivation 
of the latter
based on the
Laplace transform technique 
in the LO and NLO approximations. 
We find that the Laplace transform
method provides correct behaviors of
extracted gluon densities and
that our results 
demonstrate comparability
with parametrization groups.
Main advantage of our approach is related with simple
analytical expressions which reproduce the
essential features of more complex calculations.

{\it Acknowledgements.} 
We thank M.A.~Malyshev for the interest, very important comments and remarks. 
Phuoc Ha would like to thank Professor Loyal Durand for his invaluable support.
This research has been carried out at the expense of the Russian Science Foundation grant No. 25-22-00066, https://rscf.ru/en/project/25-22-00066/.


\vspace{1cm} 

\noindent
{\bf Appendix. Analytic results for the Sudakov form factor}

\vspace{0.5cm} 
Here we present the analytic results for the Sudakov form factor given by~(\ref{eq-sudakov}).
After some algebra, we obtain the the following expression for
$\ln T_a({\mathbf k}_T^2, \mu^2)$ with angular ordering condition, $\Delta=k/(k+\mu)$:
\begin{gather}
	\ln T_a({\mathbf k}_T^2, \mu^2) = - \frac{4C_a}{\beta_0} \left[
  s_1 R_a(x_0)+ I_a(x_0) +\sum_{i=1}^3 \delta_a^{(i)}(x_0) \right],      
	\label{Su1}
\end{gather}
where $a = g$ or $q$, $C_g = C_A$, $C_q = C_F$ and
\begin{gather}
s_1=\ln\left(\frac{a_s({\mathbf k}_T^2)}{a_s(\mu^2)}\right), \quad x_0=1-\Delta, \nonumber \\
R_q(x_0)=\ln\frac{1}{\Delta} -\frac{3}{4}x_0^2, \quad R_g(x_0)=\ln\frac{1}{\Delta} -\left(1-\frac{T_f}{2C_A}\right)x_0^2 + \frac{1}{12}\left(1-\frac{2T_f}{C_A}\right)x_0^3(4-3x_0), \nonumber \\
I_a(x_0)=-{1\over 2} \left[\ln  \frac{\mu^2}{{\mathbf k}_T^2} + \ln \left(\frac{{\mathbf k}_T^2}{\Lambda_{\rm QCD}^2} \right) \ln \left( {\ln {\mathbf k}_T^2/\Lambda_{\rm QCD}^2 \over \ln \mu^2/\Lambda_{\rm QCD}^2} \right)  \right], \quad \delta^{(1)}_a(x_0)=-2\beta_0a_s(\mu^2) j_{0}(x_0), \nonumber \\
\delta^{(2)}_a(x_0)=-4\beta^2_0 a_s^2(\mu^2) j^{(2)}_{0}(x_0), \quad \delta^{(3)}_q(x_0)=-3\beta_0a_s(\mu^2) j_{1}(x_0), \nonumber \\
\delta^{(3)}_g(x_0)=2\beta_0 a_s(\mu^2) \left[\left(1-\frac{2T_f}{C_A}\right) \left[ j_{2}(x_0)- j_{3}(x_0) \right] -\left(2-\frac{T_f}{C_A}\right)j_{1}(x_0)\right], \label{Ra} 
\end{gather}
with
\begin{gather}
j_{0}(x_0)=\ln x_1\ln x_0-\frac{\zeta_2}{2}-{\rm Li}_2(-x_1), \quad x_1=\frac{1-x_0}{x_0},\nonumber \\
j_{0}^{(2)}(x_0)=\ln^2 x_1\ln x_0+\frac{3\zeta_3}{2}+2{\rm Li}_3(-x_1) - 2\ln x_1{\rm Li}_2(-x_1),\nonumber \\
j_{1}(x_0)=\frac{1}{2}\left[x^2_0\ln x_1-\ln (2\Delta) +\frac{1}{2}-x_0\right],\nonumber \\
j_{2}(x_0)=\frac{1}{3}\left[x^3_0\ln x_1-\ln (2\Delta) +\frac{5}{8}-x_0-\frac{x_0^2}{2}\right],\nonumber \\
j_{3}(x_0)=\frac{1}{4}\left[x^4_0\ln x_1-\ln (2\Delta) +\frac{2}{3}-x_0-\frac{x_0^2}{2}-\frac{x_0^3}{3}\right].\label{ja}
\end{gather}
\noindent 
Here $\zeta_k$ are the Euler $\zeta_k$-values and 
${\rm Li}_k(x)$ are the Polylogariths (see \cite{Devoto}):
\begin{gather}
{\rm Li}_2(x)=-\int^1_0\, \frac{dz}{z}\,\ln(1-xz), \quad {\rm Li}_3(x)=\int^1_0\, \frac{dz}{z}\,\ln z\ln(1-xz).      
	\label{Li}
\end{gather}
\noindent 
In the case of the strong ordering condition, $\Delta=k/\mu$, the $\ln T_a({\mathbf k}_T^2, \mu^2)$ has the following form
\begin{gather}
	\ln T_a({\mathbf k}_T^2, \mu^2) = - \frac{4C_a}{\beta_0} \left[
  s_1 R_a(x_0)+ I_a(x_0) +\delta_a^{(3)}(x_0) \right],      
	\label{Su1a}
\end{gather}
where the corresponding functions in square brackets are the same as above~(\ref{Su1}), but $j_{i}(x_0)$ now have the form
\begin{gather}
j_{1}(x_0)=\frac{1}{2}\left[(x^2_0-1)\ln \Delta-x_0-\frac{x_0^2}{2}\right],\nonumber \\
j_{2}(x_0)=\frac{1}{3}\left[(x^3_0-1)\ln \Delta-x_0-\frac{x_0^2}{2}-\frac{x_0^3}{3}\right],\nonumber \\
j_{3}(x_0)=\frac{1}{4}\left[(x^4_0-1)\ln \Delta-x_0-\frac{x_0^2}{2}-\frac{x_0^3}{3}-\frac{x_0^4}{4}\right].
\label{jaa}
\end{gather}
\noindent
The expressions for $R_a(x_0)$ can be found \cite{TMDs-appr5}. 
Results for $R_a(x_0)$ and $I_a(x_0)$ are exact and
they are obtained without any approximations. Corrections $\delta_a^{(i)}(x_0)$
are decomposed by the coupling constant $a_s(\mu)$. First correction
$\delta_a^{(1)}(x_0)$ is more important and others are less important. Note that
all the results (\ref{Ra}), with the exception of the first and last of them, are consistent
with the so-called Casimir scaling, where the results for gluons and quarks are the
same except for the colored Casimir numbers $C_A$ and $C_F$.

\end{document}